\documentclass[twocolumn]{revtex4}

\usepackage{amssymb}
\usepackage{amsmath}
\usepackage{graphicx}
\usepackage[english]{babel}
\usepackage{amsfonts}
\usepackage[T1]{fontenc}
\usepackage[utf8]{inputenc}

\begin{document}
\title{
$K$-shell ionization of heavy hydrogen-like ions}
\author{O. Novak}
\email{email: novak-o-p@ukr.net}
\author{R. Kholodov}
\affiliation{Institute of Applied Physics, National Academy of Sciences of Ukraine, Petropavlivska str. 58, 40000 Sumy, Ukraine}

\author{A. Surzhykov}
\affiliation{Physikalisch-Technische Bundesanstalt, Bundesallee 100, 38116 Braunschweig, Germany \\
and Technische Universitat Braunschweig, Institut f\"ur Mathematische Physik, Mendelssohnstra{\ss}e 3, D-38106 Braunschweig, Germany}

\author{A. N. Artemyev}
\affiliation{Institut f\"ur Physik und CINSaT, Universit\"at Kassel, 
Heinrich-Plett-Stra{\ss}e 40, 34132 Kassel, Germany}

\author{Th. St\"ohlker}
\affiliation{Helmholtz-Institut Jena, Fr\"obelstieg 3,  07743 Jena, Germany}

\date{2018}

\begin{abstract} 
A theoretical study of the $K$-shell ionization of hydrogen-like ions, colliding with bare nuclei, is performed within the framework of the time-dependent Dirac equation.
Special emphasis is placed on the ionization probability that is investigated as a function of impact parameter, collision energy and nuclear charge.
To evaluate this probability in a wide range of collisional parameters we propose a simple analytical expression for the transition amplitude.
This expression contains three fitting parameters that are determined from the numerical calculations, based on the adiabatic approximation.
In contrast to previous studies, our analytical expression for the transition amplitude and ionization probability accounts for the full multipole expansion of the two-center potential and allows accurate description of nonsymmetric collisions of nuclei with different atomic numbers, $Z_1 \neq Z_2$.
The calculations performed for both symmetric and asymmetric collisions indicate that the ionization probability is reduced when the difference between the atomic numbers of ions increases.
\end{abstract}  
\maketitle  
  
\section{Introduction}

Studies of heavy ion collisions in storage rings have been the focus of research for a few decades.
For example, a number of experiments have been performed at the current GSI facility in Darmstadt 
\cite{%
Backe78, Mehler85, Cowan85, Kraemer89, Stoehlker93, Stoehlker00, Heinz00, Brandau03, Stoehlker03, Bednarz03, Gumberidze11, Stiebing84, Wille86, Nehler94, QEDSF%
}.
A variety of collision experiments involving heavy ions is planned at the future Facility for Antiproton and Ion Research (FAIR) \cite{FAIR, Gumberidze05, Stoehlker05, Gumberidze09}.
One of the key processes addressed in these studies is the inner-shell ionization of ions and atoms 
\cite{Molitoris86, Liesen86, Verma06, Soff78a, Soff78b}.
The analysis of this process provides an important benchmark for relativistic collision theories.
Moreover, it allows us to gain more insight into electron dynamics in the presence of strong electromagnetic fields.
Under such extreme conditions the relativistic, non-dipole and even QED effects become of paramount importance and can affect the ionization cross-sections.

However, theoretical analysis of the $K$-shell ionization of ions is a rather complicated task.
Usually it is performed based on various numerical methods \cite{Betz76, Soff79, Fillion85, Johnson88, Kullie01, Artemyev10, Tupitsyn10, Marsman11}.
However, numerical methods can involve complicated and time-consuming calculations. 
Furthermore, the dependence of the ionization cross-sections on the collisional parameters often can be hard to infer from the numerical results.

In this paper, therefore, we present a simple approach to estimate the probability of the $K$-shell ionization in heavy ion collisions.
Our approach is based on the method, proposed in Refs.~\cite{%
Liesen78, Mueller78, Mueller83, Bang79, Bang80, Liesen80, Bosch80}, 
where the corresponding matrix elements are approximated by a simple analytical expression that includes fitting parameters.
These parameters can be found by fitting the analytical matrix element to the numerical data.

In the past, the analytical approach for evaluation of the $K$-shell ionization probabilities was employed within the monopole approximation, where only the spherically symmetrical term is taken into account in the multipole expansion of the potential of two colliding nuclei.
The monopole approximation is well justified in the case of a small internuclear distance and nearly symmetrical collisions, e.g., $R \lesssim 500$~fm, $Z_1 \approx Z_2$.
Ionization takes place predominantly at small distances and, thus, the requirement of small internuclear separation holds true.
On the other hand, the role of charge asymmetry of colliding ions remained unclear up to now.
To better understand the behavior of the ionization cross-sections for $Z_1 \neq Z_2$ one needs to account for the higher-multipole terms in the expansion of the two-center potential. 
In this work, we perform this analysis and obtain analytical expressions for the ionization probability that can be used for asymmetric collisions.

The paper is organized as follows. In Sec.~\ref{sec:tcd} the solution of the time-dependent Dirac equation for the two-center Hamiltonian is reviewed.
The adiabatic approximation is used and the wave function is sought as an expansion in stationary quasimolecular orbitals with time-dependent coefficients.
The stationary wave functions of two-center potential are generated from a basis of eigenfunctions that are obtained within the monopole approximation.
In Sec.~\ref{sec:ionpb} the time-dependent transition amplitude is calculated within the framework of the first-order perturbation theory.
In Sec.~\ref{sec:fit} the approximate analytical expression of ionization probability is obtained using the parametrization of the corresponding matrix elements.
The analytical expression contains three fitting parameters.
In Sec.~\ref{sec:res} we discuss the fitting procedure and results. 
In particular, we study the $K$-shell ionization probability for various atomic numbers of colliding hydrogen-like and bare ions.
We find that the probability to ionize an electron from the ground hydrogenic state is reduced with an increase of the difference between $Z_1$ and $Z_2$.

Natural units ($\hbar = c = m = 1$) are used throughout the paper.

\section{Theoretical background}
\subsection{Solution of the two-center Dirac equation} \label{sec:tcd}
We start with the solution of the time-dependent Dirac equation for the motion of a single electron in the Coulomb field of two colliding nuclei.
The electronic wave function $\Psi_k(\vec{r},t)$ can be found as 
\begin{equation}
\label{tdde}
  i \frac{\partial \Psi_k(\vec{r},t)}{\partial t} = \hat H_{TC} \Psi_k(\vec{r},t),
\end{equation}
where $k$ is the set of quantum numbers to specify a particular state at $t = -\infty$ and $\hat H_{TC}$ is the two-center Hamiltonian which is given by 
\begin{equation}
\label{htc}
  \hat H_{TC} = (\vec{\alpha} \cdot \vec{p}) + \beta + V_{TC}(\vec{r},t).
\end{equation}
In this expression, $\vec\alpha$ and $\beta$ are the Dirac matrices and the two-center potential reads 
\begin{equation}
\label{VTC}
  V_{TC}(\vec{r},t) = -\frac{Z_1\alpha}{|\vec r - \vec R_1(t)|} - \frac{Z_2\alpha}{|\vec r - \vec R_2(t)|}
\end{equation}
with $\alpha$ being the fine-structure constant. 
Here we assume that the nuclei are moving along the classical trajectories $\vec{R}_1(t)$ and $\vec{R}_2(t)$.

In order to further simplify the time-dependent Dirac equations (\ref{tdde}) and (\ref{htc}) we need to expand the electronic wave function $\Psi_k(\vec r, t)$ into a set of basis functions.
The choice of such a set depends on the particular collision scenario under consideration.
For example, for relatively slow collisions, in which the relative ion velocity is much smaller that the electron velocity, one can use the adiabatic approximation.
Within this framework we expand the wave function $\Psi_k(\vec r, t)$ as a superposition of the stationary solutions $\Phi_i(\vec r, \vec R)$ of the two-center Dirac equation
\begin{equation}
\label{psi}
  \Psi_k(\vec r,t) = \sum \limits_{i} a_{ki}(t) e^{-i\chi_{i}(t)} \Phi_{i}(\vec r,\vec R).
\end{equation}
Here $\vec R = \vec R_1 - \vec R_2$ is an instant radius-vector between the nuclei.
For each fixed position of the nuclei, the wave functions $\Phi_i(\vec r, \vec R)$ can be obtained by solving the stationary Dirac equation
\begin{equation}
\label{SDE}
  \hat H_{TC} \Phi_i(\vec r, \vec R) = E_i \Phi_i(\vec r, \vec R).
\end{equation}
The phases $\chi_i(t)$ are defined by the eigenvalues $E_i$ of the stationary two-center Hamiltonian,
\begin{equation}
  \label{chi}
  \chi_i(t) = \int^t_0  E_i(R(t')) \:dt'.
\end{equation}
Time $t=0$ is set to be the time of the closest approach of the nuclei.

The time-dependent coefficients $a_{ki}(t)$ satisfy the initial conditions
\begin{equation}
  \label{akj}
  a_{ki}(t) \xrightarrow{ t \to -\infty } \delta_{ki}.
\end{equation}
The conditions (\ref{chi}) and (\ref{akj}) ensure that wave functions (\ref{psi}) approach the stationary state of the isolated atom at $t \to -\infty$.

In the present study the stationary wave functions $\Phi_i(\vec r, R)$ are numerically calculated using the approach developed in Ref.~\cite{McConnell12}. 
In this approach, two-center wave functions are constructed from eigenfunctions $\phi^n_{\kappa \mu}$ of the Hamiltonian in the monopole approximation,
\begin{equation}
\label{Phi}
  \Phi_{i\mu}(\vec r, R) = \sum_{\kappa = -K}^K \sum_n C^\kappa_{ni\mu}\phi_{\kappa\mu}^n(\vec r, R),
\end{equation}
where $n$, $\mu$, and $\kappa$ are the principal, magnetic, and angular quantum numbers, respectively.
The functions $\phi^n_{\kappa \mu}$ were obtained using the B-spline basis approach to solve the Dirac equation first described in Ref.~\cite{Johnson88}. 
To ensure the absence of the nonphysical spurious states the Dual Kinetically Balanced method from Ref.~\cite{Shabaev04} was applied.

The summation limit $K$ in Eq.~(\ref{Phi}) is chosen according to a desired accuracy. 
When $K$ is fixed, the first $2K+1$ terms are taken into account in the multipole expansion of the two-center potential
\begin{equation}
\label{VTCexp}
  V_{TC}(\vec r) = \sum\limits_{l=0}^{2K} V_l(r,R) P_l(\cos \theta),
\end{equation}
\begin{equation}
\label{Vl}
  V_l(r,R) = \frac{2l+1}{2} \int_0^\pi V_{TC}(\vec r)P_l(\cos\theta) \:\sin\theta \:d\theta .
\end{equation}

The calculations were performed using partial waves with $|\kappa| \leq 3$ constructed from 200 B-splines of order 8 in a box of size 300 natural units of length.
This set of parameters allows us to obtain the ionization cross sections with 1\% accuracy~\cite{McConnell12}.

The B-spline basis sets are constructed in a finite-size box which leads to the discretization of the Dirac continuum.
Thus, positive and negative continua decompose into sets of discrete states with eigenenergies $E_n > mc^2$ and $E_n < - mc^2$, respectively.
The process of $K$-shell ionization corresponds to the transition from the ground state to a discrete state with $E > mc^2$. 
Though the present paper focuses on the ionization, let us note that a similar process of excitation from an occupied energy level with $E < -mc^2$ to a vacant $1\sigma$ state is possible, which describes bound-free pair production.
In the case of small internuclear distance the energy gap for the pair production can be less than the ionization energy.
Nevertheless, the total probability of positron production is still much less than that of the ionization \cite{Nehler94, Reinhardt81}.

\subsection{Ionization probability} 
\label{sec:ionpb}

Inserting Eqs.(\ref{psi}) and (\ref{chi}) into the time-dependent Dirac equation (\ref{tdde}), one obtains the set of first-order coupled differential equations for the amplitudes $a_{kn}(t)$,
\begin{equation}
\label{cceq}
  \dot{a}_{kn}(t) = - \sum\limits_i a_{ki} e^{i(\chi_n-\chi_i)} 
  \left<\Phi_n(\vec r, \vec R)\left|\frac{\partial}{\partial t}\right|\Phi_i(\vec r, \vec R)\right>.
\end{equation}
Here, the time derivative operator acts on the states $\Phi_i(\vec r, R)$ only via their dependence on the time-dependent vector $\vec R$
and therefore can be written as
\begin{equation}
  \label{dt2dr}
  \frac{\partial}{\partial t} = \dot{\vec{R}}\!\cdot\! \frac{\partial}{\partial \vec{R}} \:.
\end{equation}

Note that at this point no approximation is involved and  the set of equations (\ref{cceq}) is mathematically equivalent to the original Dirac equation (\ref{tdde}).
However, in order to obtain the solution of Eq.~(\ref{cceq}) in a closed form, we will use an approximate approach based on the perturbation theory.
Assuming weak coupling of the solutions we set $a_{ii} \sim 1$ and $a_{ki} \ll 1$ for $k \neq i$ on the right side of Eq.~(\ref{cceq}).
The amplitude $a_{kn}(t)$ can be written then as
\begin{equation}
  \label{perturb}
  a_{kn}(t) \approx - \int\limits_{-\infty}^t dt \: e^{i(\chi_n-\chi_k)} 
  \left<\Phi_n(\vec r, \vec R)\left|\frac{\partial}{\partial t}\right|\Phi_k(\vec r, \vec R)\right>.
\end{equation}

The amplitudes $a_{kn}(t)$ describe the transition of an electron from state $k$ to state $n$ in the course of a collision. 
Thus, the amplitude for the transition from the initial bound state to a final continuum state with the energy value $E$  at $t = +\infty$ reads
\begin{equation}
  \label{anE}
  a_{k}(E) \approx - \int\limits_{-\infty}^\infty
  dt\: e^{i(\chi_E-\chi_k)}
  \left<\Phi_E\left|\frac{\partial}{\partial t}\right|\Phi_k\right>.
\end{equation}
Here the upper limit of integration is set to $+\infty$ accordingly.
Finally, the total probability of ionization from the $k$th state in a collision with given impact parameter $b$ is  
\begin{equation}
  \label{pfroma}
  P_k(b) = 2\int_0^\infty |a_k(E)|^2 dE,
\end{equation}
where the factor 2 arises due to the spin degeneracy.
Indeed, in the case of a two-center potential the angular momentum projection to the internuclear axis is conserved.
Every electronic orbital is therefore characterized by a unique eigenvalue $\mu$.
The solution with $\pm|\mu|$ is degenerate in energy since there is no difference in rotation around the internuclear axis.
Thus, in order to include electron transitions from the ground state to the final states with both angular momentum projections $+|\mu|$ and $-|\mu|$ the factor 2 is introduced in Eq.~(\ref{pfroma}).

\subsection{Analytical expression for $K$-shell ionization probability} %
\label{sec:fit}%
As seen from Eqs.~(\ref{anE}) and (\ref{pfroma}), evaluation of the ionization probability is traced back to the matrix element 
$\left<\Phi_E\left|\frac{\partial}{\partial t}\right|\Phi_k\right>$.
While in general this matrix element can be obtained only numerically, a simple analytical expression can be derived for the ionization from the ground $1\sigma$ state.
This expression reflects the main features of the matrix element
$\left<\Phi_E\left|\frac{\partial}{\partial t}\right|\Phi_k\right>$
which will be discussed below.

To start the analysis of the transition amplitude 
$\left<\Phi_E\left|\frac{\partial}{\partial t}\right|\Phi_k\right>$
we first consider in detail the operator $\partial/\partial t$. %
Since the states $\Phi_k$ are the molecular states oriented along the internuclear axis one can split the time derivative into radial and angular parts
\begin{equation}
\label{ddt}
  \frac{\partial}{\partial t} = \dot R \frac{\partial}{\partial R} - i (\vec\omega \cdot \vec j)
\end{equation}
where $\vec\omega$ is the angular velocity of the internuclear axis and $\vec{j}$ is the electron angular momentum operator.
The matrix elements of the second term in (\ref{ddt}) are known to vanish at small internuclear distances~\cite{Mueller78}.
On the other hand, the $1\sigma$ state couples to $n\sigma$ with large radial matrix elements that exhibit a very strong peak at small distances.
Thus, the second part of the operator (\ref{ddt}) can be safely neglected and the matrix element can be written as 
$\left<\Phi_E\left|\frac{\partial}{\partial t}\right|\Phi_{1\sigma}\right> \approx
\dot R \left<\Phi_E\left|\frac{\partial}{\partial R}\right|\Phi_{1\sigma}\right>$,
where $\Phi_E$ are chosen to be the $\sigma$ states with zero orbital momentum.

Our approach is based on the approximation of the radial matrix element
$ \left<\Phi_E\left|\frac{\partial}{\partial R}\right|\Phi_{1\sigma}\right> $
by an analytical function.
However, direct use of any fitting procedure is not possible due to the discretization of the continuum mentioned at the end of Sec.~\ref{sec:ionpb}.
Transitions to many of the discrete levels are strongly suppressed.
As a result, consequent matrix elements can have substantially different values. 
This problem can be solved by introduction of the averaged matrix element:
\begin{equation}
  \label{MER}
  M_{1\sigma}(E,R) = \sqrt{\frac{1}{\Delta E} \sum_n \left|
  \left< E_n \left| \frac{\partial}{\partial R} \right| 1\sigma \right>
  \right|^2}.
\end{equation}
Here the averaging interval $\Delta E$ is centered at the value~$E$. 
The number of energy levels within the averaging interval should be chosen large enough to make the averaged values smooth and suitable for parametrization with a continuous function.
Based on the numerical analysis, we set the number of levels in the interval $\Delta E$ to 25.

Note that the quantity $M_{1\sigma}(E,R)$ obtained from Eq.~(\ref{MER}) is real valued. 
Therefore, Eq.~(\ref{MER}) is applicable only if the matrix elements 
$\left<E_n|\partial/\partial R|1\sigma\right>$
are real or purely imagine.
As was previously mentioned, in the present study only transitions to $\sigma$ states with zero orbital momentum are taken into account.
This ensures that the matrix elements $\left<E_n|\partial/\partial R|1\sigma\right>$ are real valued and justify the use of Eq.~(\ref{MER}).

The important feature of $M_{1\sigma}(E,R)$ is its smooth dependence on the energy $E$ of the emitted electron.
Based on the analysis of numerical results it was shown in Refs.~\cite{Mueller78, Bosch80} that for ionization into states with $E < 3mc^2$ the approximation of the radial matrix elements can be written as
\begin{equation}
\label{fit}
  M_{1\sigma}(E,R) \approx
  \frac{\sqrt{D}}{2\pi} \:\: E^{-\frac{\gamma}{2}}   R^{-\frac{\delta}{2}},
\end{equation}
Here, $\delta$, $\gamma$, and $D$ are parameters that can be found from fitting of the function (\ref{fit}) to the numerical calculations.
It should be emphasized that in the monopole approximation all parameters in Eq.~(\ref{fit}) depend only on the combined charge of the nuclei $Z = Z_1 + Z_2$.
In contrast, in the present work the numerical values to be approximated with Eq.~(\ref{fit}) are obtained using two-center functions.
As a result, the transition amplitude depends on each charge number $Z_1$ and $Z_2$ separately.
Consequently, the parameters $D$, $\gamma$, and $\delta$ are the functions of two variables $Z_1$ and $Z_2$.
This gives the possibility to study different pairs of colliding nuclei with the same total charge $Z_1 + Z_2$ and to consider how such asymmetry affects the probability of the ionization process.

The parametrization (\ref{fit}) is not applicable when $Z_1 + Z_2 \lesssim 120$.
In particular, matrix elements vanish at small distances in this case.
Nevertheless, the inverse power law still gives a rough estimation of the general behavior of the averaged matrix element $M_{1\sigma}(E,R)$ for  sufficiently large values of the internuclear distance.

By inserting the matrix element (\ref{fit}) into Eq.~(\ref{anE}) we find the amplitude for the ionization from the ground state
\begin{eqnarray}
\label{afit}
  a_{1\sigma}(E) = \frac{\sqrt{D}}{i\pi} E^{-\frac{\gamma}{2}}
  \int\limits_{R_0}^{\infty} \frac{dR}{R^{\frac{\delta}{2}}} 
  \sin \varphi,
\\
\label{varphi}
\label{aphase}
  \varphi = \int\limits_{R_0}^{R}\frac{dR'}{\dot R(R')} (E - E_{1\sigma}).
\end{eqnarray}
Here $R_0$ is the distance of closest approach of the nuclei and $\varphi$ is the difference of the corresponding phases $\chi_j$ defined by Eq.~(\ref{chi}).

As seen from Eqs.~(\ref{afit}) and (\ref{aphase}), the knowledge of $\vec{R}(t)$ and its derivative is needed for the evaluation of the ionization amplitude $a_{1\sigma}(E)$.
As already stated before, the nuclei move along the classical trajectories $\vec{R}_1(t)$ and $\vec{R}_2(t)$.
To proceed with the ionization probability we need to choose the particular form of these trajectories.
Here we consider the case of Rutherford scattering of point-like nuclei interacting via the pure Coulomb potential.
Thus, the radial velocity $\dot R(t)$ is given by the formula
\begin{equation}
  \dot R(R) = \frac vR \sqrt{(R-R_0)(R+R_0-2a)},
\end{equation}
with
\begin{equation}
    R_0 = a + \sqrt{a^2 + b^2}, \quad
    a = \frac{Z_1Z_2 e^2}{2E_{cm}}
\end{equation}
for impact parameter $b$, center-of-mass bombarding energy $E_{cm}$ and velocity $v$ at infinity.
The phase difference (\ref{varphi}) can be rewritten as
\begin{equation}
  \varphi = 
  \frac{R_0}{v} \int_0^\rho \frac{(E-E_{1\sigma}) (1+\rho')d\rho'}{\sqrt{\rho'(\rho' + 2 - 2a/R_0)}},
\end{equation}
where $\rho = R/R_0 - 1$.
Ionization takes place mostly at small distances so that one may set $\rho \ll 1$ and estimate the integral as 
\begin{equation}
\label{inner}
  \varphi \approx w\sqrt{\rho},
\end{equation}
where
\begin{equation}
\label{omega}
  w = \frac{1}{2}k(E-E^0_{1\sigma}), \qquad  k = \frac{4R_0}{v\sqrt{2-2a/R_0}},
\end{equation}
and $E_{1\sigma}^0 = E_{1\sigma}(R_0)$. 
The actual value of the factor $w$ satisfies the condition $w \gtrsim 1$.

Inserting the phase difference (\ref{inner}) into Eq.~(\ref{afit}) and carrying out integration over $R$, the transition amplitude $a_{1\sigma}(E)$ can be written as
\begin{equation}
\label{a1sK}
  a_{1\sigma}(E) = \frac{ 2\sqrt{D} }{ i\sqrt{\pi}\Gamma(\frac{\delta}{2}) }
      E^{-\frac{\gamma}{2}}  
      \left(  \frac{2R_0}{w}  \right)^{1-\frac{\delta}{2}}
      K_{\frac{3-\delta}{2}}(w),
\end{equation}
where $K_\nu(w)$ is the  modified Bessel function of the second kind~\cite{Abramowitz}.
By making use of the amplitude (\ref{a1sK}), we can evaluate the total ionization probability (\ref{pfroma}). 
After simple algebra we find
\begin{equation}
\label{pb}
  P_{1\sigma}(b) = 2D \: \frac{ (4R_0)^{2-\delta} k^{\gamma-1} }{\pi\Gamma^2(\delta/2)} \: 
  I_{\gamma\delta}(k,E_{1\sigma}^0)
\end{equation}
where the function $I_{\gamma\delta}(k,E)$ is defined according to
\begin{equation}
\label{spec}
  I_{\gamma\delta}(k,E) = 
  \int\limits_{k}^\infty \frac{(x-kE)^{\delta-1}}{x^{\gamma}}
  \: K_{\frac{\delta-3}{2}}^2  \left( \frac{x-kE}{2} \right) \:dx.
\end{equation}

\section{Results and discussion} \label{sec:res}
Having derived the analytical expressions for the transition amplitude (\ref{afit}) and the total ionization probability (\ref{pb}) we are ready to analyze now the $K$-shell ionization in ion-ion collisions.
To perform this analysis it is convenient to introduce the total nuclear charge $Z = Z_1 + Z_2$ and the degree of charge asymmetry
\begin{equation}
  A=\frac{Z_1 - Z_2}{Z_1 + Z_2}.
\end{equation}
Obviously the quantity $A$ vanishes in symmetric collisions and always $|A| \leq 1$. 

To analyze how the asymmetry affects the ionization probability one first needs to determine the parameters $D$, $\gamma$,  and $\delta$ which enter the matrix element (\ref{fit}).
We find these parameters by fitting Eq.~(\ref{fit}) to the numerical values.
In the present work we calculate matrix elements for about 350 pairs of pointlike nuclei with combined charge ranging from $Z = $130 to 175 and asymmetry degree values in the interval from $A =$0 to 0.6.
The use of the two-center wave functions (\ref{Phi}) allows us to consider the dependence of the functions  $D$, $\gamma$,  and $\delta$ on the asymmetry degree. 
In contrast, in the monopole approximation they depend only on the total charge~$Z$.

For each set of $Z$ and $A$, values of the functions  $D$, $\gamma$, and $\delta$ have been determined by the least square fitting in the interval of the electron escape energy up to 3~MeV  and for internuclear distances from 20 to 100~fm.

It has been found that $D$, $\gamma$, and $\delta$ can be reasonably approximated by quadratic polynomials,
\begin{equation}
  \label{Dgdfit}
  \begin{array}{l}
    D = 5.50 - 1.79\zeta - 4.65A^2 \\
    \gamma = - 0.81 + 6.20\zeta - 3.44 \zeta^2, \\
    \delta = -12.08 + 20.82\zeta - 7.74 \zeta^2 + 0.34 A^2,
  \end{array}
\end{equation}
where $\zeta =\alpha Z$ and $\alpha$ is the fine-structure constant.
In this case, the typical approximation error is about 2\% and the maximum error approaches 10\% for the parameter $D$ when $Z$ is small.

\begin{figure}
  \resizebox{\columnwidth}{!}{\includegraphics{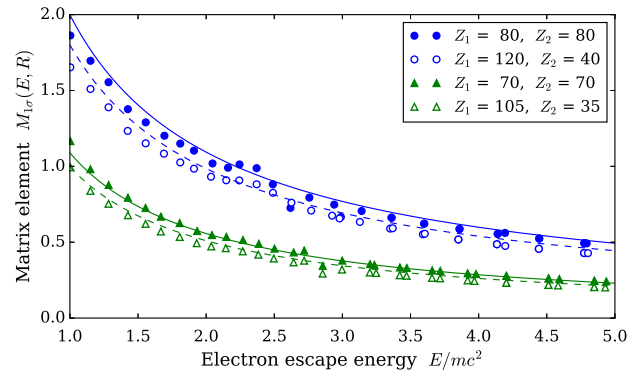}}
  \caption{%
  Averaged radial matrix elements $M_{1\sigma}(E,R)$
  (natural units) versus the electron escape energy in symmetric and asymmetric collisions with a total nuclear charge of $160$ (circles) and $140$ (triangles).
  The internuclear distance is set to 40~fm. 
  The predictions, based on the analytical expression (\ref{fit}) with fitting parameters given by Eq.~(\ref{Dgdfit}), are displayed by a solid line for symmetric and a dashed line for asymmetric collisions.
  }
  \label{fig:edep}
\end{figure}

\begin{figure}
  \resizebox{\columnwidth}{!}{\includegraphics{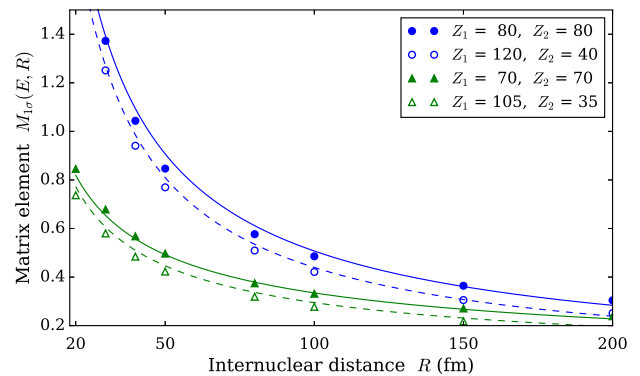}}
  \caption{%
  Same as Fig.~\ref{fig:edep} but the matrix element $M_{1\sigma}(E,R)$ is displayed as a function of the internuclear distance for the electron escape energy $E = 2mc^2$.
  }
  \label{fig:rdep}
\end{figure}

In order to illustrate how the approximate expression (\ref{fit}) can be used to reproduce the ``exact'' matrix elements we present Figs.~\ref{fig:edep} and \ref{fig:rdep}. 
Figure~\ref{fig:edep} shows the dependence of the averaged matrix elements on the electron escape energy at the internuclear distance of 40~fm and Fig.~\ref{fig:rdep} shows the dependence on the distance when the escape energy is $2mc^2$.
Calculations have been performed for symmetric, $A=0$, as well as asymmetric, $A=0.5$, collisions of ions with the total nuclear charges $Z = 140$ and $Z = 160$. 
The results of numerical computations, depicted by circles (for $Z = 160$) and triangles (for $Z = 160$), were compared with the predictions based on the analytical expression (\ref{fit}). 
These approximate results are displayed by solid and dashed lines for symmetric and asymmetric collisions, respectively.
The values of the fitting parameters $D$, $\gamma$, $\delta$ are given by Eqs.~(\ref{Dgdfit}).
Note that Eqs.~(\ref{Dgdfit}) are chosen to give the best fit of Eq.~(\ref{fit}) to the numerical data for the entire range of both escape energy and distance between nuclei and not only for the parameters used in Figs.~\ref{fig:edep} and~\ref{fig:rdep}.

\begin{figure}
  \resizebox{\columnwidth}{!}{\includegraphics{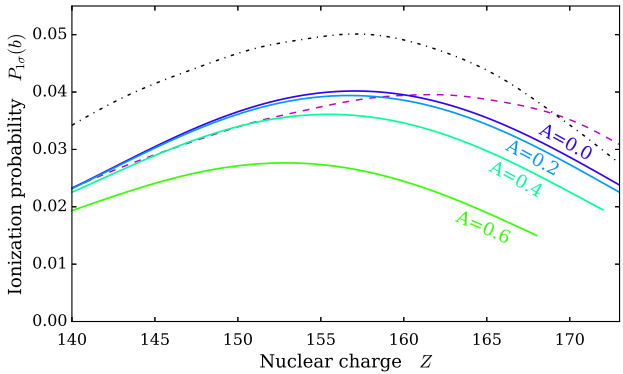}}
  \caption{
  Ionization probability of the $1\sigma$ state of a hydrogenlike ion colliding with a bare nucleus.
  The probability is calculated for a zero impact parameter and for the distance of the closest approach $R_0 = 20$~fm. 
  Solid lines show the approximation given by the formula (\ref{pb}) for different values of the asymmetry degree $A$, 
  the dashed line shows the numerical results given in Ref.~\cite{Soff78b}, 
  and the dot-dashed line shows the approximation obtained in Ref.~\cite{Mueller78} within the monopole approximation.
  }
  \label{fig:pb}
\end{figure}

The ionization probability of the $1\sigma$ state given by the approximate expression (\ref{pb})  is shown in Fig.~\ref{fig:pb}.
The impact parameter is set to zero and the distance of the closest approach of the nuclei is $R_0 = 20$~fm.
Solid lines correspond to different values of the asymmetry parameter~$A$.
The results of our calculations are compared, moreover, with the predictions of Refs.~\cite{Mueller78,Soff78b}. 
In these works, the monopole approximation was used in order to estimate the $K$-shell ionization probability in ionic collisions. 
In Fig.~\ref{fig:pb} the dashed line shows the numerical results of~\cite{Soff78b}.
Furthermore, in Refs.~\cite{Mueller78, Mueller83} an analytical approach similar to Eq.~(\ref{fit}) was employed but only with two fitting parameters involved.
The corresponding prediction for the ionization probability is depicted with the dot-dashed line.

It is clear from general considerations that in the case of symmetric collision the ionization probability should be well described by the results obtained within the monopole approximation.
Indeed, Eq.~(\ref{pb}) reproduces numerical calculations (dashed curve) quite well for $A = 0$. 
Compared to the analytical formula of Ref.~\cite{Mueller78}, the accuracy of our approach is improved due to the introduction of the additional fitting parameter~$\delta$ in Eq.~(\ref{fit}).
Moreover, our analytic approach has allowed us to take into account asymmetric collisions. 
It can be concluded the nuclear-charge asymmetry results in suppression of the ionization probability compared to the symmetric case. 
The reduction of the probability is more significant for heavy nuclei and approaches $\sim 30$\% for $A \approx 0.5$.

In order to explain the sensitivity of the ionization probability $P_{1\sigma}(b)$ to the asymmetry parameter $A$ let us consider first the function $D$, which enters the expression (\ref{fit}). 
As seen from Eq.~(\ref{Dgdfit}), this function 
is the most sensitive one to the asymmetry degree and decreases in asymmetric collisions.
It results in suppression of the ionization probability, as can be easily seen from Eq.~(\ref{pb}).

\begin{figure}
  \resizebox{\columnwidth}{!}{\includegraphics{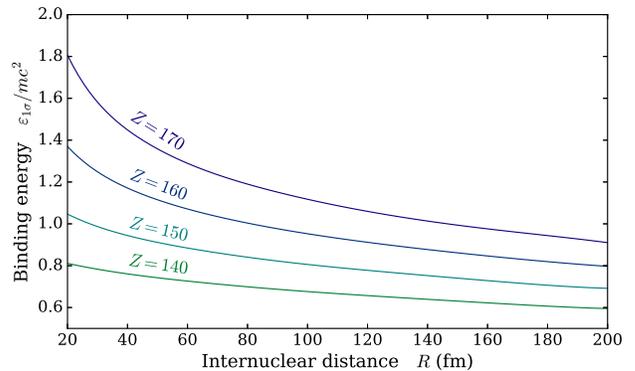}}
  \caption{%
  Binding energy of the $1\sigma$ state in symmetric collisions as a function of the internuclear distance.
  Different lines correspond to different values of the total charge $Z$.
  }
  \label{fig:e0}
\end{figure}

Another reason for the reduction of the ionization probability is the lowering of the $1\sigma$ energy level in nonsymmetric quasimolecules, which leads to a greater energy gap between the ground state and continuum states.
Based on our calculations, we found that the binding energy $\varepsilon_{1\sigma}$ of the ground state steadily grows as a square of the degree~$A$ when the total charge $Z$ and the internuclear distance are fixed,
\begin{equation}
\label{E1sfit}
    \varepsilon_{1\sigma} = \varepsilon_{1\sigma}^0  (1 + \eta A^2).
\end{equation}
Here $\varepsilon_{1\sigma}^0$ is the binding energy in the symmetric case $A=0$ with the same values of total nuclear charge $Z$ and distance $R$.
The coefficient $\eta$ depends on $Z$ and $R$ and can be approximated as
\begin{equation}
\label{koeffit}
    \eta \approx 1.70 + 1.90R - 1.42 \zeta R -4.38\zeta + 2.71\zeta^2,
\end{equation}
where $\zeta = \alpha Z$.
Figure~\ref{fig:e0} shows the dependence of the binding energy of a symmetric quasimolecule on the internuclear distance for various values of the total charge~$Z$.

The suppression of the ionization probability in asymmetric collisions is not unexpected.
Indeed, in the limit case $Z_1 \gg Z_2$ the second nucleus can be considered as a perturbation.
Reduction of the degree $A$ with fixed combined charge $Z$ results in an increase of the perturbation caused by the second nucleus, while binding of the electron to the quasimolecule weakens.
Thus, the ionization probability is maximal for symmetric collisions and decreases when the asymmetry degree $A$ grows.
  
\begin{figure}
  \resizebox{\columnwidth}{!}{\includegraphics{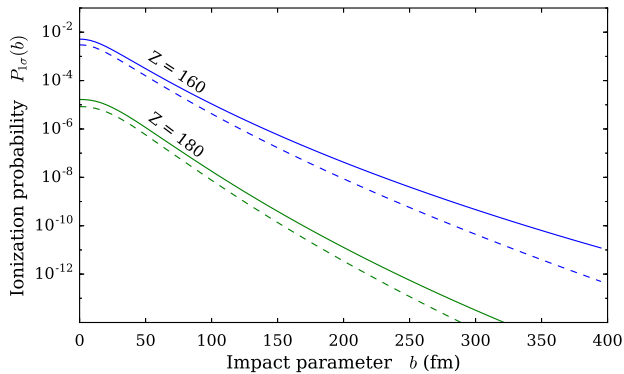}}
  \caption{%
  Impact parameter dependence of the ionization probability. Solid lines show symmetric collisions with $Z=180$ and $Z = 160$ and dashed lines collisions involving a nucleus with atomic number 118. 
  }
  \label{fig:impact}
\end{figure}

\begin{figure}
  \resizebox{\columnwidth}{!}{\includegraphics{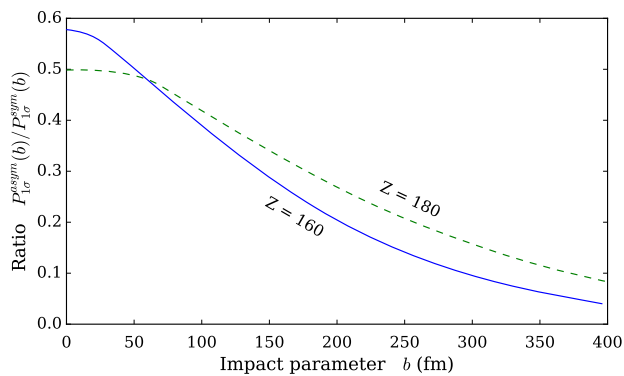}}
  \caption{
  Ratio $P_{1\sigma}^{asym}(b)/P_{1\sigma}^{sym}(b)$ of the ionization probabilities in asymmetric and symmetric collisions as a function of the impact parameter $b$. In the asymmetric case the atomic number of the heavy nucleus is set to 118.}
  \label{fig:pbratio}
\end{figure}

Figure~\ref{fig:impact} shows the dependence of the ionization probability on the impact parameter for the cases $Z = 160$ and $Z=180$. 
The center-of-mass bombarding energy is set equal to 3.5 and 2.5 MeV/nucleon, respectively.
Solid lines correspond to fully symmetric collisions (Hg+Hg and Th+Th) and dashed lines correspond to collisions involving the heaviest known element $_{118}$Og as an example.
In the considered cases the ratio of the probabilities for asymmetric and symmetric collisions $P_{asym}(b)/P_{sym}(b)$  exceeds 50\% at $b=0$ (Figure~\ref{fig:pbratio}). 

It should be stressed that reduction of the probability of  ionization in asymmetric collisions cannot be reproduced within the monopole approximation despite the small internuclear distances, since the monopole potential is not dependent on the asymmetry degree.
Thus, a nonperturbative treatment is needed for asymmetric collisions.

\section{Summary}
We presented an approximate formula for estimation of the probability of ionization from the $1\sigma$ state in collisions of hydrogenlike ions with bare nuclei.
The approach is similar to the one developed in Refs.~\cite{Mueller78, Mueller83}.
In this method, the ionization probability is obtained within the first-order perturbation theory using adiabatic expansion of the time-dependent wave function of an electron in the potential of colliding nuclei in terms of the stationary states. 
The matrix elements of transitions from the ground state to the positive continuum are parametrized with a simple expression (\ref{fit}) with three fitting parameters $D$, $\gamma$, and $\delta$.
Numerical calculations of the matrix elements were carried out using two-center wave functions obtained within the approach developed in Ref.~\cite{McConnell12}.
In contrast to the previous studies \cite{Mueller78}, our formula is based on the full multipole expansion of the two-center potential and allows one to study collisions between nuclei with different atomic numbers $Z_1 \neq Z_2$.

In the case of symmetric collisions, the values of ionization probability obtained from Eq.~(\ref{pb})  are in good  agreement with the numerical calculations carried out in the monopole approximation~\cite{Soff78b}.
In contrast, in asymmetric collisions the estimated probability shows a significant drop of the order of tens of percent. 
It can be concluded that the monopole approximation tends to overestimate the ionization probability in nonsymmetric collisions despite the small distance between nuclei.

\end{document}